# The NuMAX Long Baseline Neutrino Factory Concept[*]

J-P. Delahaye [a], C.M. Ankenbrandt [b], S.A. Bogacz [c], P. Huber [d], H.G. Kirk [e], D. Neuffer [f], M.A. Palmer [e], R. Ryne [g], and P.V. Snopok [h]

[a] *SLAC,*
 *Menlo Park, California, USA*

[b] *Muons, Inc,*
 *552 N. Batavia Avenue, Batavia, IL 60510, USA*

[c] *JLAB,*
 *Newport News, Virginia 23606, USA*

[d] *Virginia Tech.,*
 *Virginia Polytechnic Institute, Blacksburg, Virginia 24061, USA*

[e] *BNL,*
 *Upton, Long Island, New York 11973, USA*

[f] *FNAL,*
 *Batavia, Illinois 60510, USA*

[g] *LBNL,*
 *Berkeley, California 94720, USA*

[h] *IIT,*
 *Illinois Institute of Technology, Chicago, Illinois 60616, USA*

 *E-mail:* jean-pierre.delahaye@cern.ch

**ABSTRACT:** A Neutrino Factory where neutrinos of all species are produced in equal quantities by muon decay is described as a facility at the intensity frontier for exquisite precision providing ideal conditions for ultimate neutrino studies and the ideal complement to Long Baseline Facilities like LBNF at Fermilab. It is foreseen to be built in stages with progressively increasing complexity and performance, taking advantage of existing or proposed facilities at an existing laboratory like Fermilab. A tentative layout based on a recirculating linac providing opportunities for considerable saving is discussed as well as its possible evolution toward a muon collider if and when requested by Physics. Tentative parameters of the various stages are presented as well as the necessary R&D to address the technological issues and demonstrate their feasibility.

---

[*] This work was supported by Fermilab under contract No. DE-AC02-07CH11359 and by other US laboratories under contracts DE-AC02-76SF00515, DE-AC05-06OR23177, DE-SC0012704, and DE-AC02-05CH11231 with the U.S. Department of Energy.

## 1.1 Overview

The 2012 major discoveries of the large flavour mixing angle $\theta_{13}$ at Daya Bay in China and of the Higgs boson by LHC at CERN dramatically modified the Particle Physics landscape. Although the Higgs discovery corresponds to a splendid confirmation of the Standard Model (SM) and no sign of physics Beyond Standard Model (BSM) has (yet) been detected at LHC, BSM physics is necessary to address basic questions which the SM cannot, especially dark matter, dark energy, matter-antimatter asymmetry, and neutrino mass. Therefore the quest for BSM physics is a high priority for the future of High Energy Physics. It requires facilities at both the high energy and high intensity frontiers. Neutrino oscillations are irrefutable evidence for BSM physics with the potential to probe up to extremely high energies. Although neutrino studies in Long Baseline Neutrino Facilities as the one foreseen at FNAL are presently sufficient due to the unexpected large measured value of the flavour mixing angle, Neutrino Factories with an intense and well defined flux of neutrinos from muon decay will be required in the future to provide an ideal tool for high precision flavour physics at the intensity frontier. At the energy frontier, a multi-TeV lepton collider will be necessary as a precision facility to complement the LHC, for physics beyond the Standard Model if and when such physics is confirmed.

A Neutrino Factory (NF) where neutrinos are produced as tertiary particles by muons decay (Figure 1 b) would constitute the ideal complement [1] to Long Baseline Facilities based on a more standard technology where neutrinos are produced as secondary particles by pion decay (Figure 1 a), and would provide very attractive improvements especially:

- the production of all neutrino species allowing physics with multiple channels,
- a neutrino beam constitution defined with a precision of ~1% improving systematic precision
- a clean muon detector with a magnetic field to distinguish $\mu^+$ from $\mu^-$.

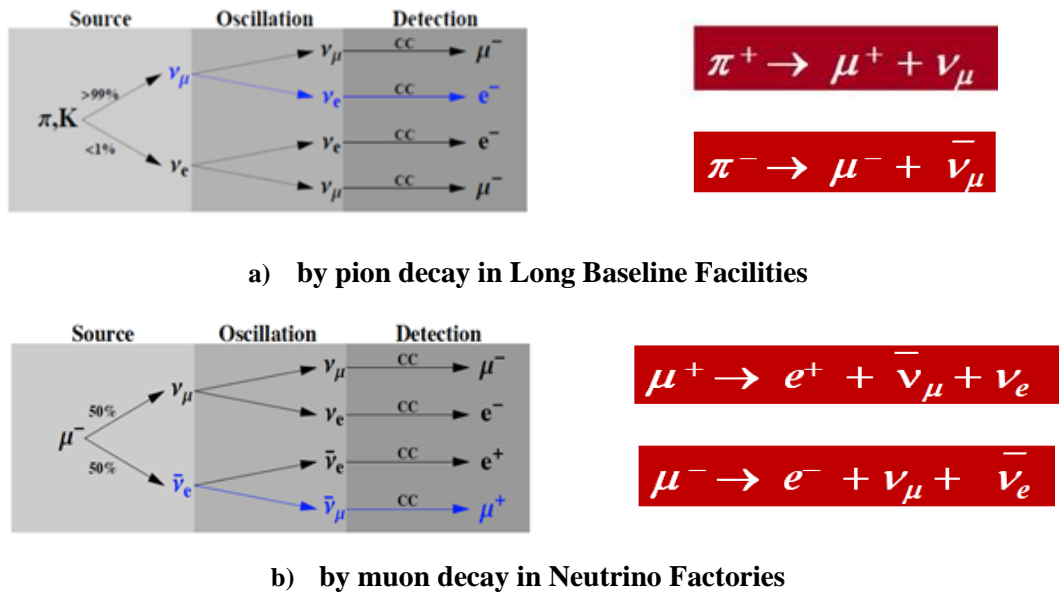

a) by pion decay in Long Baseline Facilities

b) by muon decay in Neutrino Factories

Figure 1: Neutrinos productions schemes

The concept of a Neutrino Factory on the FNAL site, called NuMAX – which stands for "**Neu**trino from **M**uon **A**ccelerator comple**X**", has been defined in the frame of the Muon Accelerator Program (MAP) [2, 3, 4]. It is strongly inspired from the IDS-NF study [5] of an ideal Neutrino Factory on a green site. Nevertheless, its concept, described in chapter 2, is significantly modified to take into account and take advantage of the specificities of the FNAL site in order to mitigate its cost and maximize the synergies with the FNAL existing or planned systems and programs.

In particular, it envisions using the Sanford Underground Research Facility (SURF) foreseen to house the Deep Underground Neutrino Experiment (DUNE) detector of the Long Baseline Neutrino Facility (LBNF). Because its distance of 1300 km from FNAL is shorter than the 2000 km considered in the IDS-NF, the optimum neutrino energy is around 1 to 2 GeV such that the muon energy is reduced from 10 to about 5 GeV with considerable savings of the accelerating system as described in the Fast Acceleration Systems chapter of [3] and the muon decay ring as described in the Neutrino Factory Storage Ring chapter of [3]. In spite of the reduced energy, the physics performance of NuMAX with a similar neutrino flux is similar to the one of IDS-NF as discussed in chapter 4

NuMAX is foreseen to be built in phases, as presented in chapter 3, in order to make the project as realistic as possible and to favor its possible future evolution towards a Muon Collider [6]. NuMAX takes advantage of the strong synergies between Neutrino Factory and Muon Collider layouts as shown on Figure 2 thus enabling facilities at both the intensity and the energy frontiers.

The R&D required to demonstrate its feasibility, optimize its performance and/or mitigate its cost is presented in chapter 5.

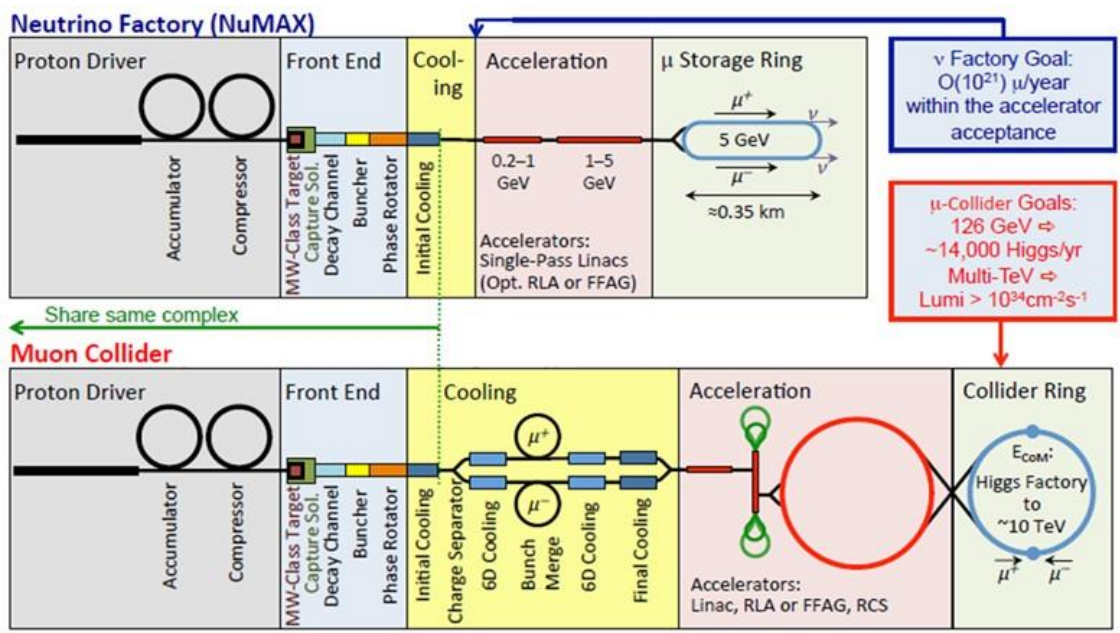

**Figure 2: Neutrino Factory and Muon Collider layouts emphasizing synergies between the various sub-systems especially concerning the muon production and initial cooling stage**

# 2 NuMAX

## 2.1 *The accelerator complex*

Like the IDS-NF [5], NuMAX uses a high-power proton beam to produce charged pions by impinging a high Z material target. The majority of the produced pions have momenta of a few hundred MeV/c, with a large momentum spread, and transverse momentum components that are comparable to their longitudinal momentum. Hence, the daughter muons are produced at low energy within a large longitudinal and transverse phase-space. This initial muon population must be confined transversely, captured longitudinally, and have its phase-space manipulated to fit within the acceptance of an accelerator. These beam manipulations must be done quickly, before the muons decay with a lifetime at rest of $\tau_0 = 2.2$ μs. Finally, muons are stored in the decay ring to produce neutrino beams in the ring's straight sections pointing towards short and long baseline detectors where neutrinos are analyzed.

The functional elements of a Neutrino Factory, illustrated schematically in Figure 2, are:
- A proton source producing a high-power multi-GeV bunched proton beam **[7]**.
- A pion production target that operates within a high-field solenoid. The solenoid confines the pions radially, guiding them into a decay channel **[8, 9]**.
- A front-end made of a solenoid decay channel equipped with RF cavities that captures the muons longitudinally into a bunch train, and then applies a time-dependent acceleration that increases the energy of the slower (low-energy) bunches and decreases the energy of the faster (high-energy) bunches **[10, 11]**.
- A cooling channel that uses ionization cooling to reduce the transverse phase space occupied by the beam, so that it fits within the acceptance of the first acceleration stage **[12, 13, 14, 15, 16]**.
- An acceleration scheme that accelerates the muons to 5 GeV **[17]**.
- A 5 GeV "racetrack" storage ring with long straight sections **[18]**.
- Short and long baseline detectors, described in the chapter 2.2

A tentative block diagram of the NuMAX complex is displayed on Figure 3

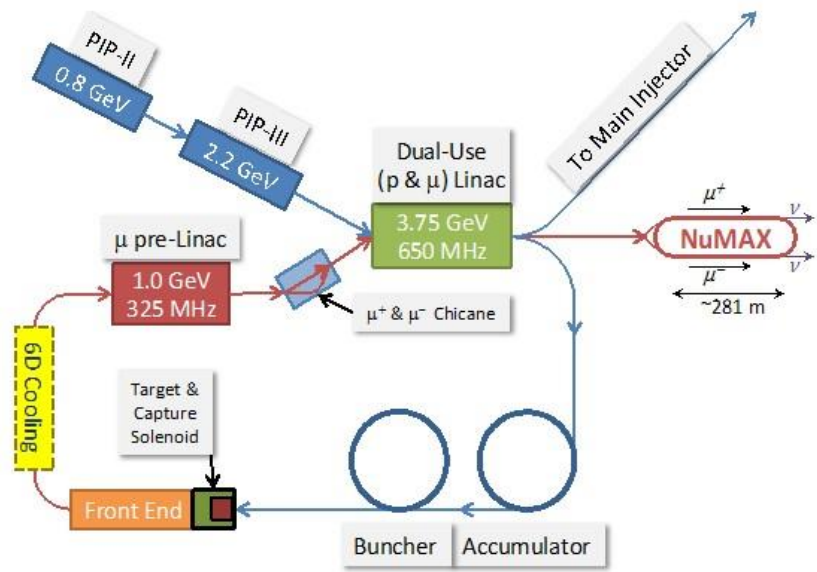

**Figure 3: Tentative block diagram of the NuMAX complex based on a dual use linac [17] accelerating both protons and muons**

It is based on an extension of the envisioned PIP-II linac accelerating the proton beam in two stages up to 3 GeV and further accelerated by a 650 MHz dual linac before hitting the target for pion production. The muons produced by pion decay, captured and bunched in the front end are recirculated to the dual linac for further acceleration up to 5 GeV as required by NuMAX. The dual use linac concept **[17]** accelerating both the proton and muon beams provides an opportunity for considerable savings as described in the Fast acceleration systems subsection of **[3]**. It requires initial cooling **[12]** to match muon beam emittances to the linac acceptances at the 325 and 650 MHz RF standards adopted by the PIP-II program. The initial cooling specifications result from a cost optimization as the best trade-off between linac, RF and cooling.

## *2.2 The detector*

With a baseline of 1300 km corresponding to the distance between Fermilab and the Sanford Underground Research Facility (SURF), the relevant neutrino energies for oscillation measurements (dictated by $\Delta m_{32}^2$) lie in the 1–2 GeV range. The MIND technology preferred in the International Design Study for the Neutrino Factory (IDS-NF) starts to become inefficient at these low energies and it is anticipated that a change of detector technology will be needed. Two candidates suggest themselves at this point in time: magnetized, fully active, plastic scintillator and magnetized liquid argon TPCs. Since the DUNE detector **[19]** of the LBNF facility has chosen a liquid argon (LAr) TPC as its far-detector technology, a staged approach to a Neutrino Factory using a magnetized liquid argon detector seems the way to go, with possibly 10 kt fiducial mass (twice as much for the whole detector) at the initial phase of NuMAX, upgradable to 30 kt at the final phase of NuMAX+.

A 12-meter diameter by 60-meter long liquid argon detector requires a coil with 400,000 A/m to generate a 0.5 T magnetic field. One might use NbTi carrying 40,000 A in a 10-cm diameter cryostat. This would require 23,000 meters of cable or 920,000 kA-m. The flux return would require 14 kilotons of steel.

There is considerable liquid argon TPC R&D taking place worldwide with the primary goal of providing input to the detailed design of the DUNE far detector(s). There have been some R&D efforts in Europe toward a magnetized LAr TPC, but considerable R&D remains to be done. Pending that R&D, it is not yet clear whether a non-magnetized LAr TPC for DUNE could be economically retrofitted with a magnetic field or whether an entirely new detector would need to be built.

With such detector, the physics performance **[1]** compares well with the performance of other facilities even in the early NuMAX phases as shown in Figure 4. The gradual upgrade of the facility and the detector allows progressive improvements in the performance of the facility towards the required precision of a few degrees in the CP violating phase.

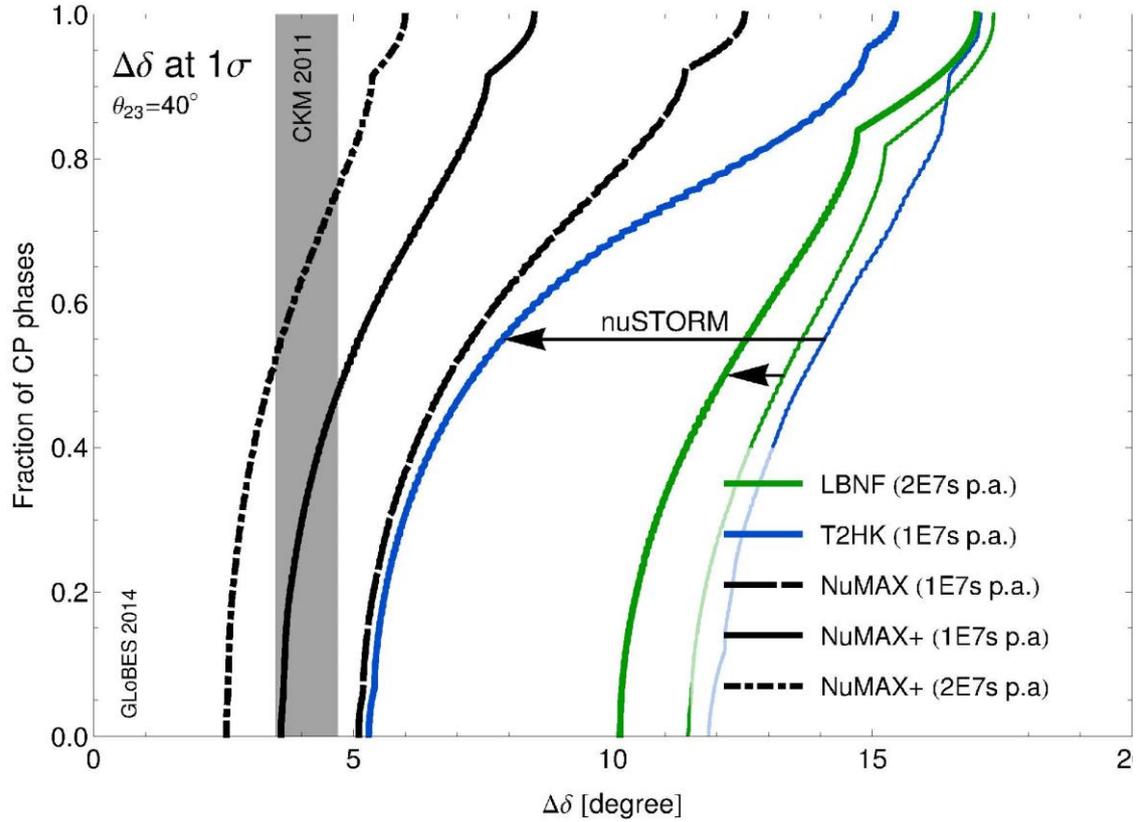

**Figure 4**: NuMAX stages Physics performance [1] in terms of CP phase δ compared with the one of Long Baseline Neutrino facilities like LBNF and T2HK and their possible improvement by precise measurements of neutrino cross sections at nuSTORM [20]

# 3 Phased approach

## 3.1 Rationale

The feasibility of the technologies required for Neutrino Factories and/or Muon Colliders must be validated before a facility based upon these can be proposed. Such validation is usually made in dedicated test facilities, which are specially designed to address the major issues. Although very convenient, these test facilities are rather expensive to build and to operate over several years. They are therefore difficult to justify and fund, given especially that they are usually useful only for technology development rather than for physics.

An alternative approach is considered here. It consists of a series of facilities built in stages, where each stage offers:
- Unique physics capabilities such that the corresponding facility obtains support and is funded.
- An integrated R&D program, in addition to the science program, that supports technology development, beam tests, validation of subsequent steps, and the acquisition of operational experience for the next stage.
- Construction of each stage as an add-on to the previous stages, extensively reusing the equipment and systems already installed, such that the additional budget of each stage remains affordable.

## 3.2 Staging scenario

A complete staging scenario has been developed within the framework of the Muon Accelerator Staging Study (MASS) [21]. It consists of a series of facilities, each with increasing complexity and with performance characteristics providing unique physics reach:

- **nuSTORM** [22] : a short-baseline Neutrino Factory-like ring enabling a definitive search for sterile neutrinos, as well as neutrino cross-section measurements that will ultimately be required for precision measurements at any long-baseline experiment.
- **NuMAX (N**eutrino from **M**uon **A**ccelerator Comple**X):** a long-baseline 5 GeV Neutrino Factory, optimized for a detector at the **S**anford **U**nderground **R**esearch **F**acility (SURF). This facility can be deployed in phases:
    - An **initial (commissioning) phase** based on a limited proton beam power of 1MW on the muon production target with no cooling for an early and realistic start with conventional technology, while already providing very attractive physics parameters.
    - **NuMAX baseline** upgraded from the commissioning phase by adding a limited amount of 6D cooling, affording a precise and well-characterized neutrino source that exceeds the capabilities of conventional superbeams.
    - **NuMAX+**, a full-intensity Neutrino Factory, achieved through progressive upgrades of NuMAX. These upgrades include increasing the proton beam power on target, when it becomes available, and carrying out a corresponding upgrade of the detector for performance similar to that specified for the IDS-NF [5]. Such a machine represents the ultimate source to enable precision CP-violation measurements in the neutrino sector.
- **Higgs Factory** [23, 24]: a collider capable of providing between 3500 (at startup) and 13,500 Higgs events per year ($10^7$ sec) with exquisite energy resolution enabling direct Higgs mass and width measurements.
    - Possible upgrade to a **Top Factory** with production of up to 60000 top particles per year ($10^7$ sec) for precise top properties measurements.
- **Multi-TeV Collider**: if warranted by LHC results, a muon collider with an ultimate energy reach of 10 TeV likely offers the best performance, lowest cost, and minimum power consumption of any lepton collider capable of operating in the multi-TeV regime.

Such a staging scenario provides clear decision points before embarking upon each subsequent stage. This represents an especially attractive approach at FNAL, where it leverages the use of current and planned facilities, thus maximizing the synergies between the ongoing FNAL program and the proposed MAP path forward, specifically:
- Existing tunnels and other conventional facilities;
- The Proton Improvement Plan (PIP) as the MW-class proton driver for muon generation;
- The Sanford Underground Research facility (SURF) as developed for the DUNE detector, which could then house the detector for a long-baseline Neutrino Factory.

Obviously, some parts of the plan could be skipped depending on Physics needs.

A tentative block diagram of the overall complex in a phased approach emphasizing the evolution and synergies from Neutrino Factory to Muon collider is shown in Annex II. The systems installed for each phase is re-used in the following phases for which specific equipment or sub-systems are added after tests and validation in the previous phase.

# 4   Main parameters

Preliminary parameters of the three NuMAX phases with progressively increasing complexity and performance are presented in Annex 1 and compared with nuSTORM. In particular, the final phase of NuMAX+ provides a neutrino flux similar to the one obtained by IDS-NF **[5]**.

In order to achieve the required flux of $5 \times 10^{20}$ neutrinos per year at the far detector, 60 bunches of $3.5 \times 10^{10}$ muons/bunch are stored in the muon decay ring with a 15 Hz repetition rate. Taking into account reasonable transmission performance, including muon decay losses, along the NuMAX complex and a production rate of 0.13 useful muon per 6.75 GeV proton on target **[10]**, it requires a high but not unreasonable proton beam power of 2.75 MW on target for muon production and a modest amount of 6D cooling by a factor 50 (5 in each transverse plane and 2 in the longitudinal direction) in order to match the muon beam emittances to the acceptances of a reasonable accelerating system.

The early NuMAX commissioning phase without any cooling and a proton beam power of 1 MW, corresponding to the present state of the art, already provides an attractive flux although one order of magnitude lower than the one provided by the IDS-NF. The flux is then improved by about a factor 4 by implementing the 6D cooling in the NuMAX baseline phase.

# 5   R&D

Since the initial-stage Neutrino Factory, NuMAX, relies on a proton beam power of 1 MW without any cooling, its critical challenges are limited to:
- Proton driver and target corresponding to the state of the art in operation at SNS and therefore no specific development needed, except possibly for the dual use linac accelerating both protons and muons beams;
- A 15–20 T solenoid to efficiently capture the pions produced in the target. One would try to limit the magnetic field to ~ 10-12 T so that $Nb_3Sn$ conductor could be used;
- Accelerating gradient in low frequency (325–975 MHz) RF structures immersed in high magnetic field as required by the front end;
- High efficiency Recirculating Linear Accelerators (RLA);
- 10 kt magnetized liquid argon (LAr) or magnetized fully active plastic-scintillator detector.

The required high-field solenoid and RF cavities immersed in large magnetic fields have been major subjects of development during the MAP Feasibility Assessment phase. The novel RLA technology involves multi-pass arcs based on linear combined-function magnets, which allow two consecutive passes with very different energies to be transported through the same string of magnets. Such a solution combines compactness with all of the advantages of a linear non-scaling FFAG, namely, the large dynamic aperture and momentum acceptance essential for large-emittance muon beams. The dogbone RLA with 2-pass arcs is the subject of a specific proof-of-concept electron test facility, JEMMRLA (JLab Electron Model of Muon RLA [25], proposed to be built and operated at Jefferson Lab. The NuMAX facility could thus be built soon after the completion of the R&D specified in the MAP feasibility study.

The baseline Neutrino Factory, NuMAX, is upgraded from the NuMAX initial stage by modest 6D cooling of the beam emittances by a factor 2 in longitudinal and 5 in both transverse planes. Its major technical challenge therefore consists of:
- Ionization Cooling, which is being studied in the MICE experiment at RAL, with first results expected in 2018. As described in Ref. [20], ionization cooling at reasonable intensity ($10^8$ muons/bunch) could be further tested using the proposed nuSTORM facility as a muon source. In parallel, cooling at full Muon Collider intensities (~$10^{12}$ muons/bunch) could be tested with protons in the proposed ASTA test facility at FNAL.

The full-intensity Neutrino Factory, NuMAX+, is upgraded from the NuMAX baseline by additional proton beam power up to 2.75 MW on target. Its major technical challenges therefore consist of:
- A multi-megawatt Proton driver which is being validated by the European Spallation Source (ESS).
- A corresponding upgrade of the target possibly by adopting Hg-jet target technology from which the feasibility has successfully been demonstrated by the MERIT experiment [9] at CERN.

The NuMAX facility could then be constructed in phases with progressing complexity and performance, building up on the technical systems and the operational experience accumulated during the previous phases.

In parallel, NuMAX could be used to validate the technology required for the following phases:

- A lower-intensity Neutrino Factory, the initial NuMAX, that does not require any cooling but could be used as a long-baseline neutrino source and an R&D platform to test and validate the longitudinal and transverse cooling (6D) at full muon intensity ($10^{12}$/pulse) as required by the full-intensity Neutrino Factory, NuMAX+. In addition, it would validate the injector complex at the 1 MW level as well as the corresponding target, front end and 5 GeV RLA.

- A high-intensity Neutrino Factory, NuMAX+, only requires a modest amount of 6D cooling (by a factor 2 in longitudinal and 5 in each transverse planes) but could be used as a muon source and an R&D platform to test and validate the demanding transverse and longitudinal (6D) cooling by five orders of magnitude to full specification and nominal muon bunch intensity ($10^{12}$/bunch) as required by Muon Colliders.

# Annex I: Main beam and machine parameters of the various stages of the NuMAX Neutrino Factory compared with the preliminary nuSTORM facility

| System | Parameters | Unit | nuSTORM | NuMAX Initial | NuMAX Baseline | NuMAX+ Upgrade |
|---|---|---|---|---|---|---|
| Performance | $\nu_e$ or $\nu_\mu$ to detectors/year | - | $3\times10^{17}$ | $4.9\times10^{19}$ | $1.8\times10^{20}$ | $5.0\times10^{20}$ |
| Performance | Stored $\mu^+$ or $\mu^-$/year | - | $8\times10^{17}$ | $1.25\times10^{20}$ | $4.65\times10^{20}$ | $1.3\times10^{21}$ |
| Detector | *Far Detector:* | Type | SuperBIND | MIND / Mag LAr | MIND / Mag LAr | MIND / Mag LAr |
| Detector | Distance from Ring | km | 1.9 | 1300 | 1300 | 1300 |
| Detector | Mass | kT | 1.3 | 100 / 30 | 100 / 30 | 100 / 30 |
| Detector | Magnetic Field | T | 2 | 0.5-2 | 0.5-2 | 0.5-2 |
| Detector | *Near Detector:* | Type | SuperBIND | Suite | Suite | Suite |
| Detector | Distance from Ring | m | 50 | 100 | 100 | 100 |
| Detector | Mass | kT | 0.1 | 1 | 1 | 2.7 |
| Detector | Magnetic Field | T | Yes | Yes | Yes | Yes |
| Neutrino Ring | Ring Momentum ($P_\mu$) | GeV/c | 3.8 | 5 | 5 | 5 |
| Neutrino Ring | Circumference (C) | m | 480 | 737 | 737 | 737 |
| Neutrino Ring | Straight section | m | 184 | 281 | 281 | 281 |
| Neutrino Ring | Number of bunches | - | | 60 | 60 | 60 |
| Neutrino Ring | Charge per bunch | $1\times10^9$ | | 6.9 | 26 | 35 |
| Acceleration | Initial Momentum | GeV/c | - | 0.25 | 0.25 | 0.25 |
| Acceleration | Single-pass Linacs | GeV/c | - | 1.0, 3.75 | 1.0, 3.75 | 1.0, 3.75 |
| Acceleration | | MHz | - | 325, 650 | 325, 650 | 325, 650 |
| Acceleration | Repetition Frequency | Hz | - | 30 | 30 | 60 |
| Cooling | Hor.*Vert.*Long. | | No | No | 5*5*2 | 5*5*2 |
| Proton Driver | Proton Beam Power | MW | 0.2 | 1 | 1 | 2.75 |
| Proton Driver | Proton Beam Energy | GeV | 120 | 6.75 | 6.75 | 6.75 |
| Proton Driver | Protons/year | $1\times10^{21}$ | 0.1 | 9.2 | 9.2 | 25.4 |
| Proton Driver | Repetition Frequency | Hz | 0.75 | 15 | 15 | 15 |

**Annex II: Evolution in stages of the muon complex from a Neutrino Factory (a) to a HIGGS factory (b) and a multi-TeV Muon Collider (c)**

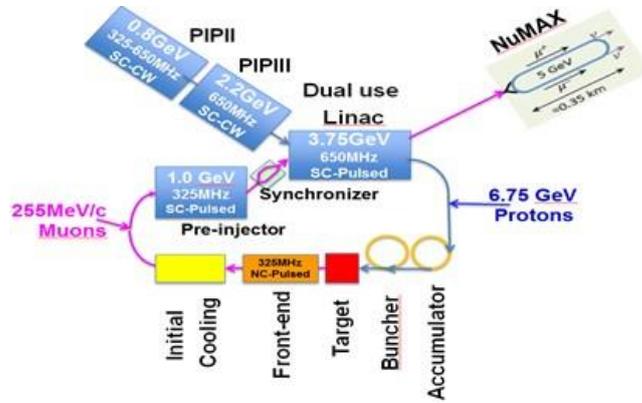

a) Layout of a Muon based Neutrino factory

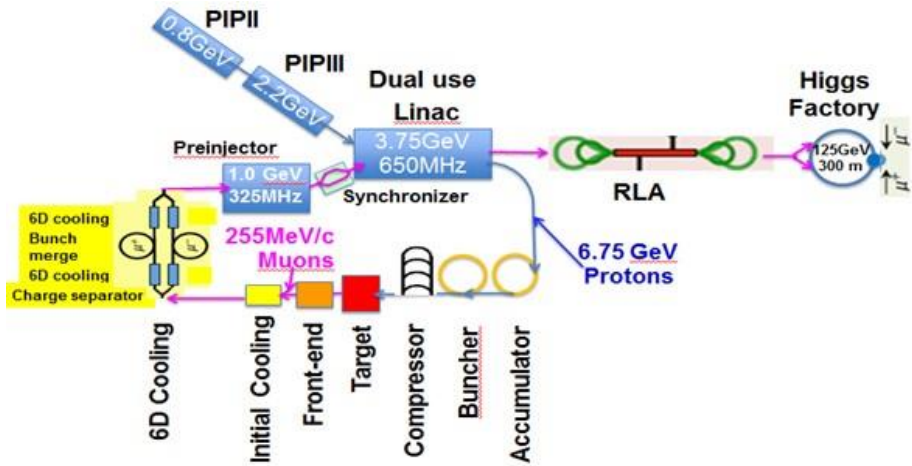

b) Layout of a Muon based Higgs factory

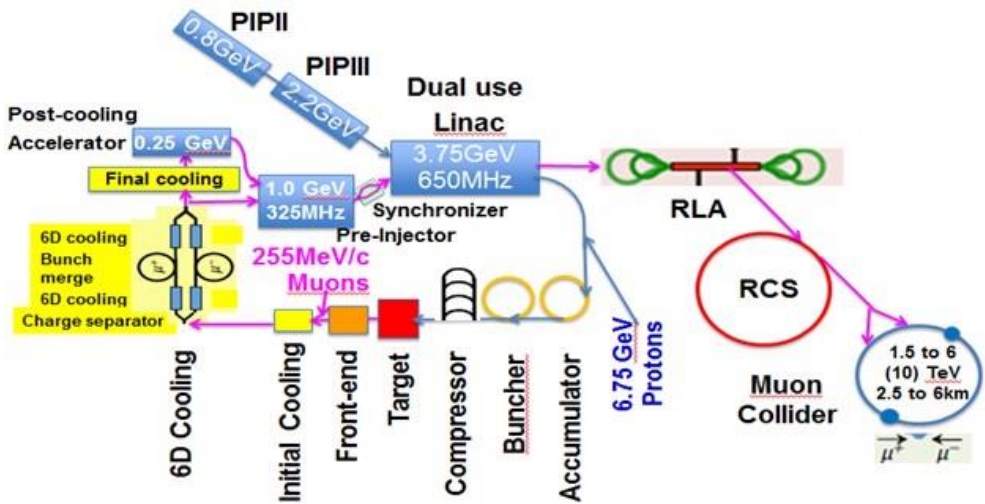

c) Layout of a multi-TeV Muon Collider